\documentstyle[12pt,procsla,epsfig]{article}
  
  \catcode`\@=11
  \long\def\@makefntext#1{
  \protect\noindent \hbox to 3.2pt {\hskip-.9pt  
  $^{{\ninerm\@thefnmark}}$\hfil}#1\hfill}		
  
  \def\@makefnmark{\hbox to 0pt{$^{\@thefnmark}$\hss}}  
  	
  \def\ps@myheadings{\let\@mkboth\@gobbletwo
  \def\@oddhead{\hbox{}
  \rightmark\hfil\ninerm\thepage}   
  \def\@oddfoot{}\def\@evenhead{\ninerm\thepage\hfil
  \leftmark\hbox{}}\def\@evenfoot{}
  \def\sectionmark##1{}\def\subsectionmark##1{}}
  
\begin{document}
  
  \centerline{\normalsize\bf NEUTRINO INDUCED EVENTS} 
  \baselineskip=16pt
  \centerline{\normalsize\bf IN THE PIERRE AUGER DETECTOR}

  \vspace*{0.6cm}
  \centerline{\footnotesize GONZALO PARENTE and \underline{ENRIQUE ZAS}}
  \baselineskip=13pt
  \centerline{\footnotesize\it Departamento de Part\'\i culas, Universidad de
Santiago} 
  \baselineskip=12pt
  \centerline{\footnotesize\it 15706 Santiago de Compostela, Spain}
  \centerline{\footnotesize E-mail: zas@gaes.usc.es}
  \vspace*{0.9cm}
  \abstracts{We study the potential of the Pierre Auger detector for
horizontal air showers initiated by ultra high energy neutrino. Assuming some 
simple trigger requirements we obtain measurable  event rates for neutrino
fluxes from AGN, from topological defects and from the interactions of cosmic
rays with the microwave background. }
   
  \normalsize\baselineskip=15pt
  \setcounter{footnote}{0}
  \renewcommand{\thefootnote}{\alph{footnote}}
  \section{Introduction}

It has been known for a long time that deeply penetrating high energy 
particles such as muons and neutrinos initiate horizontal air showers  that can
be detected at ground level \cite{berez}. Since the interaction length for
muons and 
neutrinos in the atmosphere is larger than the whole atmospheric depth, 
they have roughly equal probability to interact at any point in the 
atmosphere.  On the other hand, the rate of air showers due to the hadronic
particles, that constitute the bulk of the cosmic rays, decreases very rapidly
with zenith angle as the atmospheric depth rises from about $1000~g\,cm^{-2}$ in
the  vertical direction to close to $36000~g\,cm^{-2}$ horizontally. The
electromagnetic component of air showers started by electrons, photons and
hadrons gets absorbed well before reaching the Earth's surface and only the
muon component of the shower survives for sufficiently large zenith angles. A
detector that is able to identify the electromagnetic component of air
showers is then capable of identifying horizontal showers induced by such 
penetrating particles. Such an array will
mainly trigger on horizontal showers that initiate at the appropriate  depth 
so that the shower is close to shower maximum when it reaches the array. 

The recent agreement between calculations of diffuse neutrino fluxes from
Active Galactic Nuclei (AGN) has raised a lot of expectations for neutrino
telescopes that are currently under development and construction. 
Although horizontal showers have ruled out an early prediction
of neutrino fluxes from  AGN \cite{hawaii}, these fluxes 
extend to the PeV region where the corresponding horizontal shower rate is very
close to that expected from hard bremsstrahlung of the conventional 
atmospheric muon flux (produced in $\pi$ and $\kappa$ decays). Horizontal
showers are currently being studied by several ground arrays because they
should provide complementary information on prompt muon and neutrino production
in the atmosphere which can be related to production of charm \cite{gonzalez}
and of  cosmic ray composition around the knee \cite{gonzalo}. It is accepted
that the most appropriate technique for neutrino detection consists on 
detecting the \v Cerenkov light from muons or showers produced by the neutrino
interactions in water or ice\cite{physrep}.  

The situation is however different for still higher energy neutrinos where the
project to build two 3000~km$^2$ particle arrays one in each hemisphere
(Pierre Auger Detector) may  play an interesting role. The project is discussed
in a separate article in these proceedings\cite{Auger}. The reference design
combines an array of particle detectors and an air fluorescence device similar
to Fly's Eye to detect cosmic ray air showers of energies above  $10^{19}~$eV.
The proposed particle detectors are water \v Cerenkov tanks, very 
appropriate for
detecting particles arriving horizontally.  The detector will be most efficient
for high zenith angle showers of energy above $10^{19}~$eV when  a large number
of detectors of the array register significant signals. The electromagnetic
component is separated from the muon component on the basis of the individual
muon pulses that stand out of the average signal produced by the
electromagnetic component of the shower. Neutrino predictions of such energies 
include those from interactions of the cosmic rays with the cosmic microwave
background which have a solid foundation and would be of enormous value to
establish the Greisen-Zatsepin-Kuz'min cutoff, as well as more speculative
sources such as topological defects\cite{battarch,yoshida} and primordial black
holes\cite{macGibbon}. 

\section{Large Showers from Neutrinos}

Neutrinos produce showers in most interactions with the atmosphere but the
showers are of different nature depending on the process in consideration. In
the interactions the target nucleons break up and the debris behaves as a
group of hadrons that results in a shower similar to those induced by regular 
hadronic cosmic rays. Such shower is produced in both neutral and charged
current interactions. If the neutrino is of electron flavor there will be an 
additional shower produced by the electron at the leptonic vertex of
charged current interactions. This shower is of electromagnetic type, somewhat 
narrower and  with a smaller muon content than the hadronic showers. The
resulting shower for the charged current interactions of electron neutrinos is
thus a superposition of two parallel showers one of hadronic type and the other
electromagnetic. For high energy neutrino interactions the
average fractional energy transfer to the nucleon in the lab frame ($y$) is
$<y>\simeq 0.2$ so the electromagnetic shower carries on average $80\%$
of the neutrino energy and is therefore most important. The interactions of the
neutrinos with the electrons have in general much smaller cross sections and
can be disregarded except for the resonant electron-antineutrino electron
interaction which dominates just for neutrino energies around the resonant value
of 6.4~PeV. In that case the character of the shower depends on the
disintegration channel of the produced $W$ boson in the $s$-channel. 

The showers regardless of their character can be detected by a particle array
if they are initiated at an  appropriate distance by a neutrino with sufficient
energy. They will resemble ordinary air showers the main  difference being that 
horizontal showers develop in a more uniform atmosphere. Electromagnetic
showers should have lateral and depth distributions according to standard 
parametrizations when the corresponding lengths are measured in depth
(g~cm$^{-2}$). Similarly we assume that hadronic showers will be not too
different to ordinary cosmic ray showers. Large zenith angle showers of
energies  above the array threshold can be detected if a suitable trigger is
selected, provided the plane of the array intersects the shower at a location
where the number of particles is close to its maximum. This is a conservative 
statement since the particle detectors of the array are more closely
distributed in
the transverse plane to a near horizontal shower. The triggering to be
developed will be however very different from the standard trigger for vertical
showers, in particular the  relative timing of the signals in adjacent
detectors will reflect the shower propagation across the array. 

\section{Event Rates}

For a neutrino flux $d\Phi_{\nu}/dE_{\nu}$  interacting through a
process with differential cross section $d\sigma/dy$, where $y$ is the fraction
of the incident particle energy transferred to the target, the event
rate for horizontal showers can be obtained by a simple convolution:
\begin{eqnarray}
\Phi_{sh} [E_{sh}>E_{th}] = N_a \rho_{air} 
\int_{E_{th}}^{\infty} dE_{sh} \int_{0}^{1} dy \, \frac{d \Phi_{\nu}} 
{dE_{\nu}} (E_{\nu}) \, \frac{d \sigma}{dy} (E_{\nu},y) \, \cal{A} 
(y, E_{\nu}) 
\label{eq:13}
\end{eqnarray} 
where $N_a$ is Avogadro's number and $\rho_{air}$ is the air density. The energy
integral corresponds to the shower energy $E_{sh}$ which is related to the
primary neutrino energy $E_{\nu}$ in a different way depending on the
interaction  being considered. 
$\cal{A}$ is a geometric acceptance which
contains the volume and solid angle integrals 
for different shower
positions and orientations with respect to the array. 

\subsection{Acceptance} 

We define the effective acceptance $\cal{A}$ as the integral over volume and
solid angle in $d\Omega= d\phi d (\sin \bar{\theta})$ where $\phi$ is the
azimuthal angle in the array plane. 
It depends on the energy transfer to the shower and on the type of
shower produced in the  interaction. We take the shower axis to go through
the array and assume showers are large enough to trigger when they start at an
adequate point. The effective area is then simply given by $S \sin
\bar{\theta}$ where $S$ is the area covered by the array and
$\bar{\theta}=90^o-\theta_z$ is the angle between the neutrino arrival
direction and the array plane, (complementary to the zenith angle). To obtain
the effective volume it must be multiplied by a ''depth interval''. For a 
given neutrino direction and impact parameter the depth interval is basically
the range of positions of the interaction point that will trigger the array. As
a given shower is moved through all possible first interactions points it spans
an infinite cylinder which we refer as a shower-tube. Such tube intersects the
array plane in an ellipse with a major axis given by  $q=2 r/sin \bar{\theta}$
where $r$ is the radius of the shower-tube. We can take the projection of this
length onto the shower axis as the depth interval, which is equivalent to
demanding that the shower maximum intercepts the array. This is conservative
because it ignores the shower length which increases 
the range of allowed positions for the first interaction point. Since the
depth interval cannot exceed the length  of the array, $W$, we take the minimum
of $q$ and the average length of the array $\widehat{W}$.  

We calculate $\cal{A}$ integrating this volume over the possible solid angle
orientations of the shower, $d \Omega = 2 \pi d sin \bar{\theta}$, and restrict
the integration to horizontal showers i.e. $0^o < \bar{\theta} <
\bar{\theta}_{max} \simeq 20^o$.   
\begin{eqnarray}  \cal{A} = S \times 2 \pi
\,  \left(   \int_{\sin \bar{\theta}_1}^{\sin \bar{\theta}_{max}} \,  d(\sin
\bar{\theta} ) \sin \bar{\theta}  \frac{2r}{\sin \bar{\theta} } +
\int_{0}^{\sin \bar{\theta}_1} \,  d(\sin \bar{\theta} ) \sin \bar{\theta}~ W
\right)   \nonumber \\ 
\mbox{ }  = S \times 2 \pi \,  r (2 \sin
\bar{\theta}_{max} - \sin \bar{\theta}_1)   \label{eq:1} \end{eqnarray}  
For $\bar{\theta} < \bar{\theta}_1=\sin^{-1}(2r/\widehat{W})$, the intersection
of  the tube reaches $\widehat{W}$, its maximum value. 

\subsection{Sensitivity to Neutrino Fluxes}

It is now a matter of  substituting reasonable values for the parameters to get
an estimate for the acceptance. For $S=3000~km^2$ the ''diameter'' of the array
is approximately $D = 65~km$. For $\widehat{W}$ we should take the average
length across the array for all possible impact parameters, we obtain 
$\widehat{W} \simeq 0.70 D \sim 45~km$. The
acceptance scales with the tube radius which we take as $r=1.5~km$, the
separation between the individual detectors of the array. Ordinary cosmic ray
showers are expected to give measurable signals in detectors that are this
distance away from the shower axis. The showers we consider here have shower
maximum intercepting the array plane so they should have similar particle 
densities. We  can now obtain $\bar{\theta}_1 = 0.07$~rad and an acceptance of
$\cal{A} = 17000~km^3~sr$ which when multiplied by an air density $\rho_{air}
\simeq 1.1~10^{-3}~g~cm^{-3}$ gives $2~10^7~kT~sr$.  Neutrino detectors in
planning aim towards an active volume in the range of $1~km^3$ \cite{1km3}.
Their effective volume is enhanced because of the long range of the energetic
muon produced, but for electron and  tau neutrinos they have to collect the
the \v Cerenkov light from the showers they produce. If their energy is well
above the PeV region the Earth will be opaque to these neutrinos and the
corresponding acceptance of a $1~km^3$ detector for contained events is at most 
$6~10^6~kT~sr$, illustrating how the Pierre Auger project may come into play. 

In order to obtain a rough estimate of the rate of horizontal showers above
$10^{19}~eV$ produced by a given neutrino flux we can simply take the product
of the neutrino flux above $10^{19}$~eV, the total cross section $\sigma$ and
the acceptance $\cal{A}$. For charged current neutrino electron interactions
all the neutrino energy is transferred to the shower. The cross section
corresponding to charged current neutrino interactions at this energy is
uncertain because of the unknown behavior of the structure functions at low $x$
and high $Q^2$ which take part in the calculation. Extrapolations of the
structure functions lead to cross sections in the $\sigma=1.3-4~10^{-32}~cm^2$
range\cite{reno}. If we take the extreme neutrino fluxes from topological
defect models \cite{battarch} divided by a factor of 2 to account for electron
neutrinos, the integral neutrino flux ranges from 
$\Phi=10^{-16}~[cm^2~s~sr]^{-1}$ for the model with $p=1.5$ to 
$\Phi=4~10^{-14}~[cm^2~s~sr]^{-1}$ for the model with $p=0$. We obtain 
$2-5~10^{-8}~s^{-1}$, in the range of one event per year for the lowest flux
and about 400 times that for the highest. The result is extremely encouraging
because the calculation is very conservative. There are  several issues that
will rise the event rate: we have ignored the neutral current interactions and
muon neutrinos all together, the cross section and the acceptance integral
should both increase with energy, large
showers which are very horizontal may trigger the array even if their axis
falls outside the array area and it is also possible that the particle arrays 
have a lower threshold for horizontal showers. 
 
  \begin{figure}[hbt]
  \centering
  \vspace*{-2.5cm}
         \mbox{\epsfig{figure=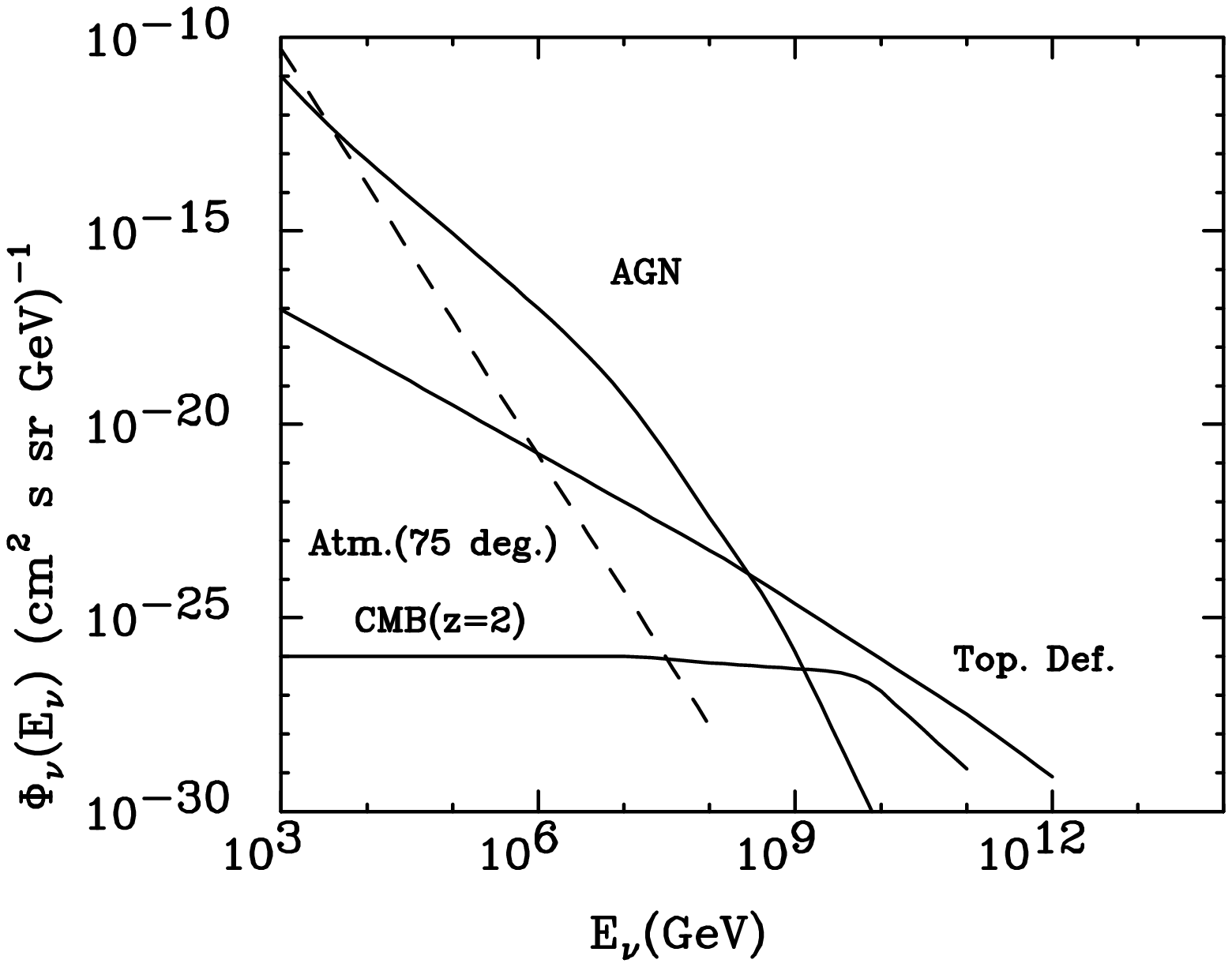,width=12.0cm}}
  \vspace*{-3cm}  
  \fcaption{Neutrino flux predictions in the EeV range.}
  \label{fig:radk}
\end{figure}

The neutrino flux predictions for topological defects have been normalized in 
a maximal way, assuming  the observed highest energy cosmic ray spectrum is due
to the topological defects themselves but it may be that such fluxes are close
to ten orders of magnitude below \cite{kibble}. It should be stressed that
there are solid predictions for neutrinos produced  in the cosmic ray
interactions with the cosmic microwave background responsible for the GZK
cutoff. In the range $10^{19}-10^{20}~eV$ they only differ from topological
defects by less than one order of magnitude (see Fig. 1). It is well possible
that such interesting events become accessible to the two detector arrays
planned.

\subsection{Full Event Rate Calculation}

For more realistic calculations, which are in progress, triggering details
become important. We estimate such effects demanding that a number of 
consecutive particle detectors in a row have an electron density above a fixed 
value. The integral for the  effective acceptance can be calculated numerically
using parametrizations of the lateral distribution functions for both
electromagnetic and hadronic showers. The results obtained in both cases are
quite similar. As the threshold electron density is decreased the detector 
increases its acceptance for showers of lower energy because these showers are
small and have to be extremely well aligned with the detector rows in order to
trigger. However for the larger energy showers the acceptance does not change
much as the threshold is lowered.  Preliminary results for three triggering
requirements are illustrated in Fig.~(2) reflecting the stability of the
results for shower energies above $10^{19}~eV$.

  \begin{figure}[hbt]
  \centering
  \vspace*{-2.5cm}
         \mbox{\epsfig{figure=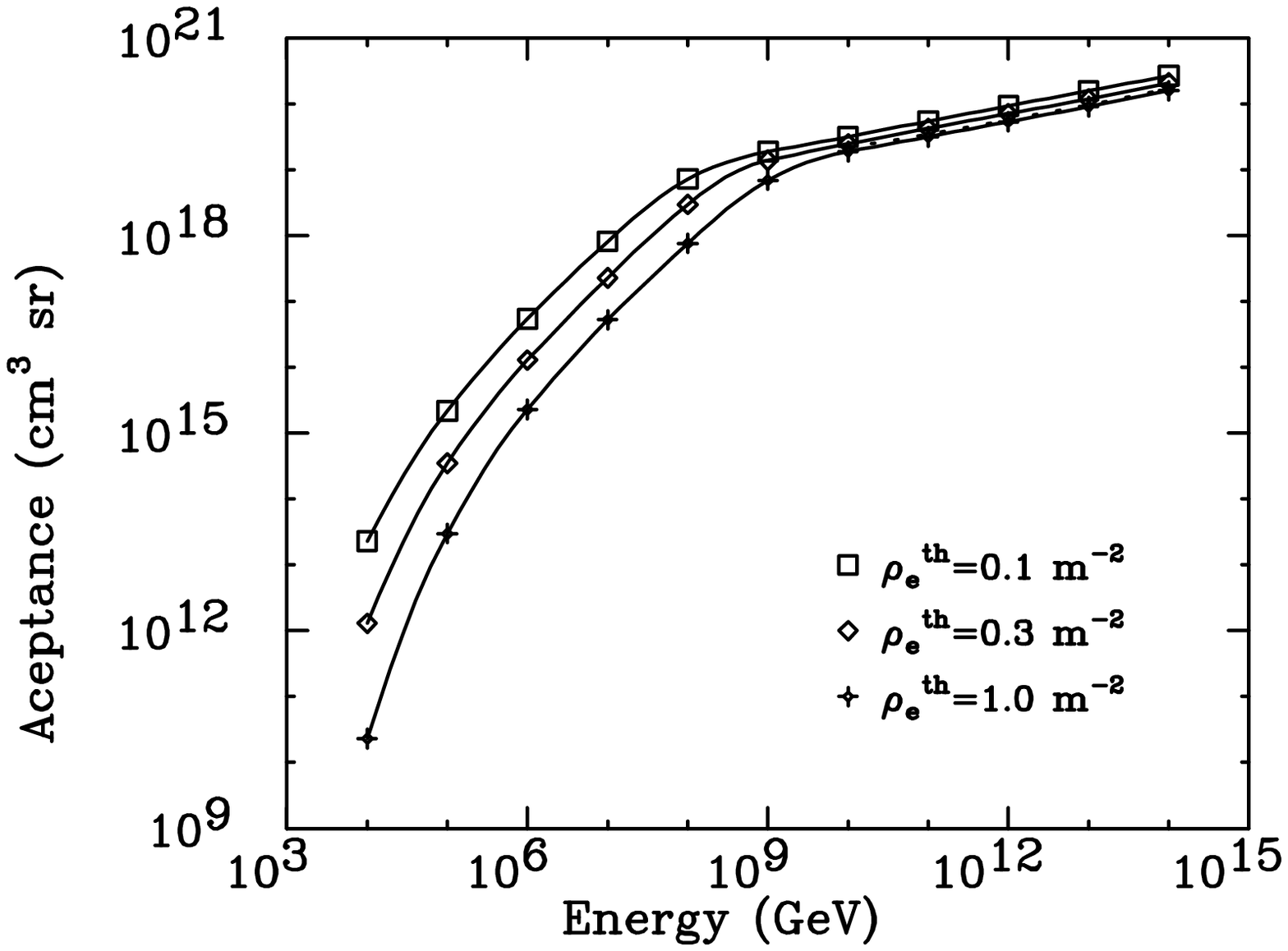,width=12.0cm}}
  \vspace*{-3.cm}
  \fcaption{Acceptance calculation for three trigger models} 
  \label{fig:fig2}
  \end{figure}

To calculate event rates we use two sets of structure functions 
MRS(G) \cite{MRSG} and  GRV \cite{GRV}. For the first we extrapolate
to low $x$ beyond validity of the parametrization using the slope of $xq(x)$,
where $q(x)$ is the standard parton distribution. The second set, GRV, can be
cautiously used  on its own for low x. Fig.~(3) shows both predictions in
comparison with a data point obtained from H1 collaboration in HERA
\cite{HERA}.  We take three neutrino fluxes for reference calculation. We use
the lowest prediction of ref. \cite{battarch,yoshida} for neutrinos produced in
the decay of topological defects with $p=1.5$. This flux is very flat and would
dominate the neutrino sky for energies above $E_{\nu}=10^{17}~eV$. We take the
upper limit of the band calculated in ref. \cite{protheroe} and the prediction of
the neutrino fluxes from cosmic ray interactions with the cosmic microwave
background calculated in ref. \cite{nuGZK} integrated up to redshift $z=2$. 
Fig.~(1)
illustrates  these fluxes compared to other predictions setting the scale of
the sensitivity of the Pierre Auger project to high energy neutrino fluxes. We
approximate the  electron neutrino flux to be a factor of two below the muon
neutrino for all three cases. This ratio can be naively  expected from the
number of channels in the decays of pions. The results are shown in table~1.

  \begin{figure}[hbt]
  \centering
  \vspace*{-2.5cm}
         \mbox{\epsfig{figure=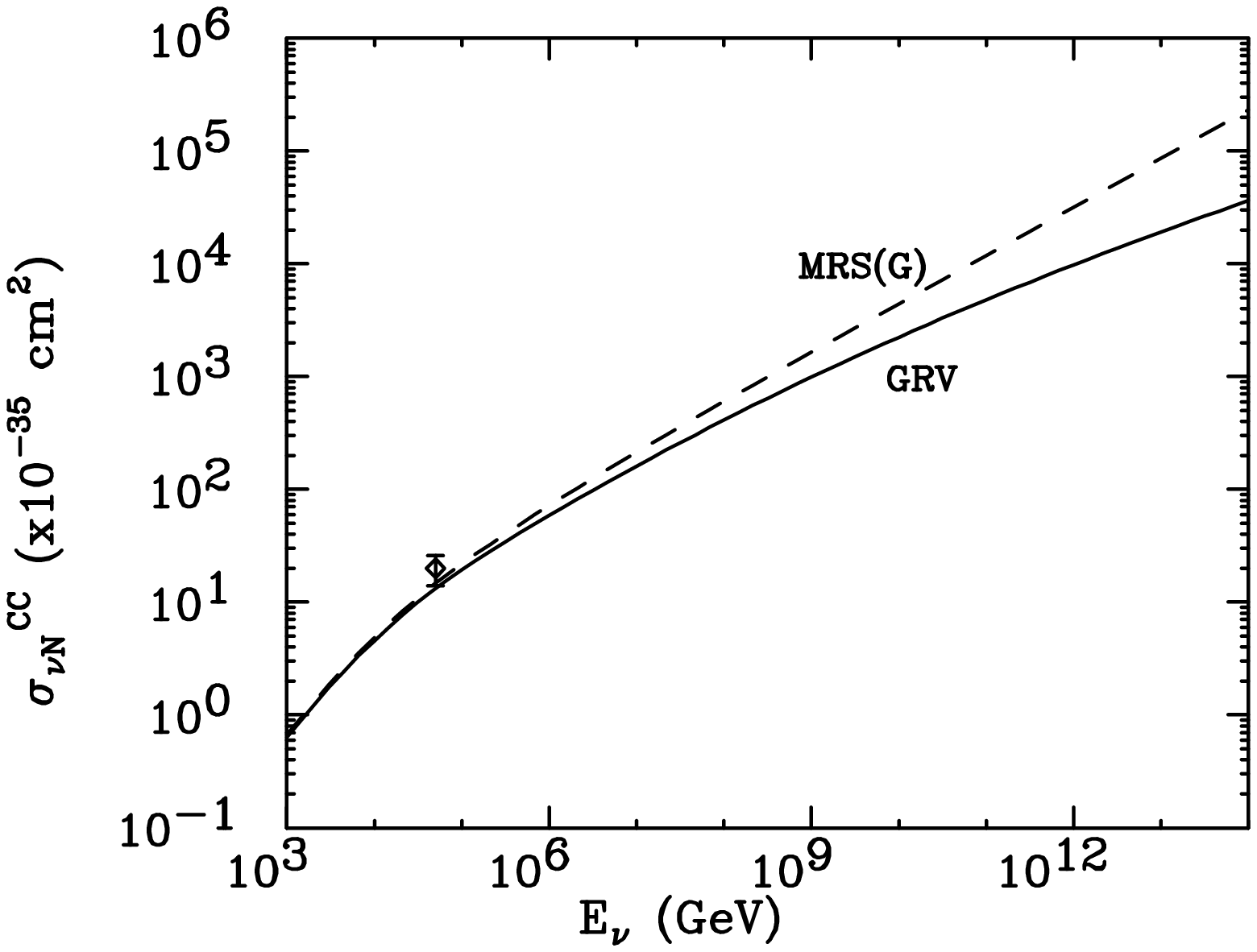,width=12.0cm}} 
  \vspace*{-3.cm}
  \fcaption{Comparison of two neutrino cross section predictions in the EeV
range}
  \label{fig:fig3}
  \end{figure}

\begin{table}[hbt]
\label{ratenu}
\begin{displaymath}
\begin{array}{||c||c|c||}
\hline
\hline
      & {\rm MRS(G)} & {\rm GRV} \\ 
\rho_e^{th} \; (m^{-2}) & & \\
\hline
& \multicolumn{2}{c||} {AGN}\\
\hline
  1 &  2 &  2 \\
0.1 & 40 & 30 \\
\hline
& \multicolumn{2}{c||} {\rm CMB}\\
\hline
  1 & 0.9 & 0.5 \\
0.1 & 2.9 & 0.9 \\
\hline
& \multicolumn{2}{c||} {\rm TD~p=1.5} \\
\hline
  1 & 26 & 9 \\
0.1 & 51 & 17 \\
\hline
\hline
\end{array}
\end{displaymath}
\caption{Yearly neutrino event rates for diffuse fluxes from AGN, for neutrinos
from the interactions with the cosmic microwave background (CMB) and for
topological defects in the model described in the text (TD).}
\end{table}

The acceptance curve shown in Fig.~(2) is a continuous function of shower 
energy and when it is combined with the AGN flux 
prediction it can give measurable rates. Because these fluxes are typically of
PeV energies they will produce small showers compared to the typical showers
detected in the Auger detector and the showers will have to run well aligned
with a row of particle detectors to trigger. These showers will undoubtedly
produce signals that are very different from those of typical showers.  
The low energy part of these curves is very sensitive to the triggering
conditions and there are large differences between event rates for different
trigger models. This is not the case for the topological defect fluxes and for
the flux from interactions of cosmic rays with the cosmic microwave background.
These fluxes are much  flatter and hence the horizontal shower rate peaks for
neutrino energies in the region where the acceptance integral is fairly stable
strengthening the results obtained. 

\section{Conclusions}

The Pierre Auger project can be made sensitive to ultra high energy neutrino
fluxes through horizontal showers if an appropriate trigger is implemented. Its
acceptance for detecting contained neutrinos events of energy above $E_{\nu}
\sim 10^{19}~eV$ will be of the order of other neutrino telescopes in planning.
The peak of horizontal shower acceptance for the Pierre Auger Project is at
energies about $10^{19}~eV$, a lot higher than the optimal region for AGN
neutrino detection, for which the conventional approach to detect neutrinos is
best suited. The Pierre Auger Project is best suited for detection of 
neutrinos from interactions of cosmic rays with the cosmic microwave background
and from the decay of topological defects. The event rates expected under
some simplifying models for the trigger are high enough to be observed.

\newpage
\noindent {\bf Acknowledgments}
\vspace{0.3cm}

We thank James W. Cronin for a large number of stimulating discussions and 
support to pursue this idea, Jaime Alvarez Mu\~niz, Jose Juan Blanco  Pillado,
Francis Halzen, Antoine Letessier-Selvon and R.~V\'azquez for  discussions and
suggestions.  This work was supported in part by the Xunta de Galicia under
contract XUGA-20604A93 and in part by and the CICYT under contract AEN93-0729.

\vspace{0.8cm}
\noindent {\bf References} 
\vspace{0.3cm}


\begin{thebibliography}{999}
%
%
\bibitem{berez}
V.S.~Berezinsky and A.Yu.~Smirnov, {\sl Astrophys.\ Space
Science\ {\bf 32} (1975) 461}.
%
\bibitem{hawaii} F. Halzen and E. Zas, Phys. Lett. B289 (1992) 184.
%
\bibitem{gonzalez} M.C. Gonz\'alez-Garc\'\i a, F.~Halzen, R.A.~V\'azquez
and E.~Zas, Phys. Rev. D49 (1994) 2310.
%
\bibitem{physrep} T.K. Gaisser, F. Halzen and T. Stanev, Phys. Rep.
238 (1995) 173.
%
\bibitem{gonzalo} G. Parente, A. Shoup and G.B. Yodh,
Astropart. Phys. 3 (1995) 17.
%
\bibitem{Auger} M.~Boratav these proceedings.
%
\bibitem{battarch}
P.~Battacharjee, C.T.~Hill, D.N.~Schramm, Phys.\ Rev.\ Lett.\ {\bf 69} (1992) 
567.                                      
%
\bibitem{yoshida} S. Yoshida et al., 
Proc. 24th ICRC, Rome, vol. 1, p. 793 (1995).
%
\bibitem{macGibbon} J.H.~MacGibbon and B.J.~Carr, Astrophys. J. {\bf 371}
(1991), 447; F.~Halzen, B.~Keszthelyi and E.~Zas, Phys. Rev. {\bf D52} (1995)
2310. 
%
\bibitem{1km3} F.~Halzen these proceedings.
%
\bibitem{reno} G.~Parente and E.~Zas, Proc.of Int. Europhys. Conf. Brussels, 
(1995); R.~Ghandi {\it et al}, Preprint FERMILAB-PUB-95/221-T (1995).
%
\bibitem{kibble}
A.J.~Gill and T.W.B.~Kibble, Phys.\ Rev.\ {\bf D50} (1994) 3660.  
%
%
\bibitem{MRSG}
A.D.~Martin, W.J.~Stirling and R.G.~Roberts,
{\sl Phys.\ Lett.\ {\bf B354} (1995) 155}. 
%
\bibitem{GRV}
M.~Gl\"uck, E.~Reya and A.~Vogt, 
{\sl Z Phys.\ C\ Particles\ and Fields\ {\bf 67} (1995) 433}. 
%
\bibitem{HERA}
T.~Ahmed et. al, H1 Collaboration, {\bf B324}{1994}{241}
%
\bibitem{protheroe}
A.P.Szabo, and R.J. Protheroe, in Proc. High Energy Neutrino Astrophysics
Workshop (U. Hawaii, March 1992), ed. V.J. Stenger, J.L. Learned, S.Pakvasa and
X. Tata, World Scientific, 1993, p. 24.
%
\bibitem{nuGZK}  F.W.~Stecker, C.~Done, M.H.~Salomon and P.~Sommers,
Phys.\ Rev.\ Lett.\ {\bf 66} (1991) 2697.
\end{thebibliography}
\end{document}